\documentclass[aps,prl,twocolumn,showkeys,superscriptaddress,floatfix]{revtex4-2}
\usepackage[T1]{fontenc}
\usepackage[utf8]{inputenc}
\usepackage{amsbsy}
\usepackage{amstext}
\usepackage{graphicx}
\usepackage[nice]{nicefrac}
\usepackage{xfrac}

\makeatletter

\usepackage[english]{babel}
\usepackage{amsmath}
\usepackage{amsfonts}
\usepackage{color}
\usepackage{amssymb}
\usepackage{array}
\usepackage{makeidx}
\usepackage{float}
\usepackage{multirow}
\usepackage{times}
\DeclareTextSymbol{\degre}{OT1}{23}

\newcommand{\VS}{V$_5$S$_8$}
\newcommand{\CVS}{(Cu$_{2/3}$V$_{1/3}$)V$_2$S$_4$}

\begin{document}


\title{Evidence of a Kondo lattice quantum critical point and of non-Fermi liquid behavior in the intercalated layered system \VS{}}




\author{Hancheng Yang}
\affiliation{IMPMC, Sorbonne Universit\'e, CNRS, MNHN, 4 place Jussieu, 75005 Paris, France}
\affiliation{Department of Physics, Zhejiang University, Hangzhou, 310058, China}
\author{Hicham Moutaabbid}
\author{Beno\^it Baptiste}
\affiliation{IMPMC, Sorbonne Universit\'e, CNRS, MNHN, 4 place Jussieu, 75005 Paris, France}
\author{David Hrabovsky}
\affiliation{Center of Low-Temperature Physical Measurements, Sorbonne Universit\'e, 4 place Jussieu 75005 Paris, France}
\author{Andrea Gauzzi}
\author{Yannick Klein}
\affiliation{IMPMC, Sorbonne Universit\'e, CNRS, MNHN, 4 place Jussieu, 75005 Paris, France}
\email[]{yannick.klein@sorbonne-universite.fr}

\date{\today}

\begin{abstract}
By means of a specific heat, susceptibility and high-pressure electrical resistivity study, we show that the local magnetic moments of the intercalated V ions in \VS{} realize a prototype of Kondo lattice system, where an antiferromagnetic order of the moments coexists with a Fermi liquid in the VS$_2$ layers with intermediate heavy Fermion properties. The antiferromagnetic order and the Fermi-liquid behavior are simultaneously suppressed at a critical pressure, $P_c =10$ GPa, signature of a quantum critical point, which supports a Kondo lattice scenario and raises the question whether, in the paramagnetic phase at higher pressures, the heavy quasiparticles survive or form a non-Fermi liquid phase governed by the Kondo interaction.   
\end{abstract}


\maketitle

Several intermetallic compounds containing $f$-electrons, known as heavy fermions (HFs), display a dramatic enhancement of the effective mass, $m^{\ast}$, of the conduction electrons, up to $\sim 10^3$ times the free-electron mass \cite{kon64}, as a result of the Kondo exchange interaction, $J$, between the spin of the conduction electrons and the local magnetic moment of the $f$-electrons \cite{ste84}. Signature of HF behavior is an anomalously large density of states at the Fermi level, $D$, typically probed by measuring the Sommerfeld electronic specific heat coefficient, $\gamma$, or the Pauli susceptibility, $\chi_P$.

Since the discovery of HF behavior in CeAl$_3$ \cite{and75}, HFs have attracted interest owing to their remarkable properties. At low temperatures, the conduction electrons behave as conventional Fermi liquid (FL) quasiparticles in spite of the heavy mass \cite{sch05}. At a characteristic temperature, $T_K$, exponentially small in the coupling constant $h=JD$, such heavy FL (HFL) phase becomes unstable towards the hybridization of the conduction electrons with the local moment. If the local moments form a (Kondo) lattice, they tend to order - usually antiferromagnetically (AFM) - below $T_N \sim h^2$, owing to the Ruderman-Kittel-Kasuya-Yoshida (RKKY) exchange interaction mediated by the conduction electrons. As schematically described by the Doniach phase diagram \cite{don77}, the competition between RKKY and Kondo interactions generates a quantum critical point (QCP) at $h_c$, separating the AFM phase at $h<h_c$ from the paramagnetic (PM) HFL phase at $h>h_c$. Prototypical Kondo lattice systems, such as CeCu$_6$ or CeRu$_2$Ge$_2$, do indeed exhibit a transition between these two phases as a function of pressure \cite{sue99}, chemical composition \cite{loe94,fri09} or magnetic field \cite{tro00}, thus confirming a QCP scenario \cite{ste01,col05,loe07,geg08}. 

Whether a similar phenomenology may occur using less localized $d$-, instead of $f$-electrons, is an open question \cite{kru03,nie23}. While some $d$-electron systems exhibit clear HF properties, the mass enhancement is less pronounced than in $f$-electron systems and the mechanism remains controversial. For the most representative compound, LiV$_2$O$_4$ \cite{ani99,hop02}, and other transition metal (TM) oxides, such as CaCu$_3$Ir$_4$O$_{12}$ \cite{mey14} and LaCu$_3$Ru$_4$O$_{12}$ \cite{rie16}, a Kondo interaction between conduction and localized $d$-electrons has been invoked. Though, some authors argued that the Hund or superexchange interactions compete with the Kondo interaction and proposed alternative mechanisms, e.g. magnetic frustration \cite{ful01} or the proximity to a correlated insulator \cite{ari04,joe07}.

\begin{figure}[ht]
	\centering 
	\includegraphics[width = \columnwidth]{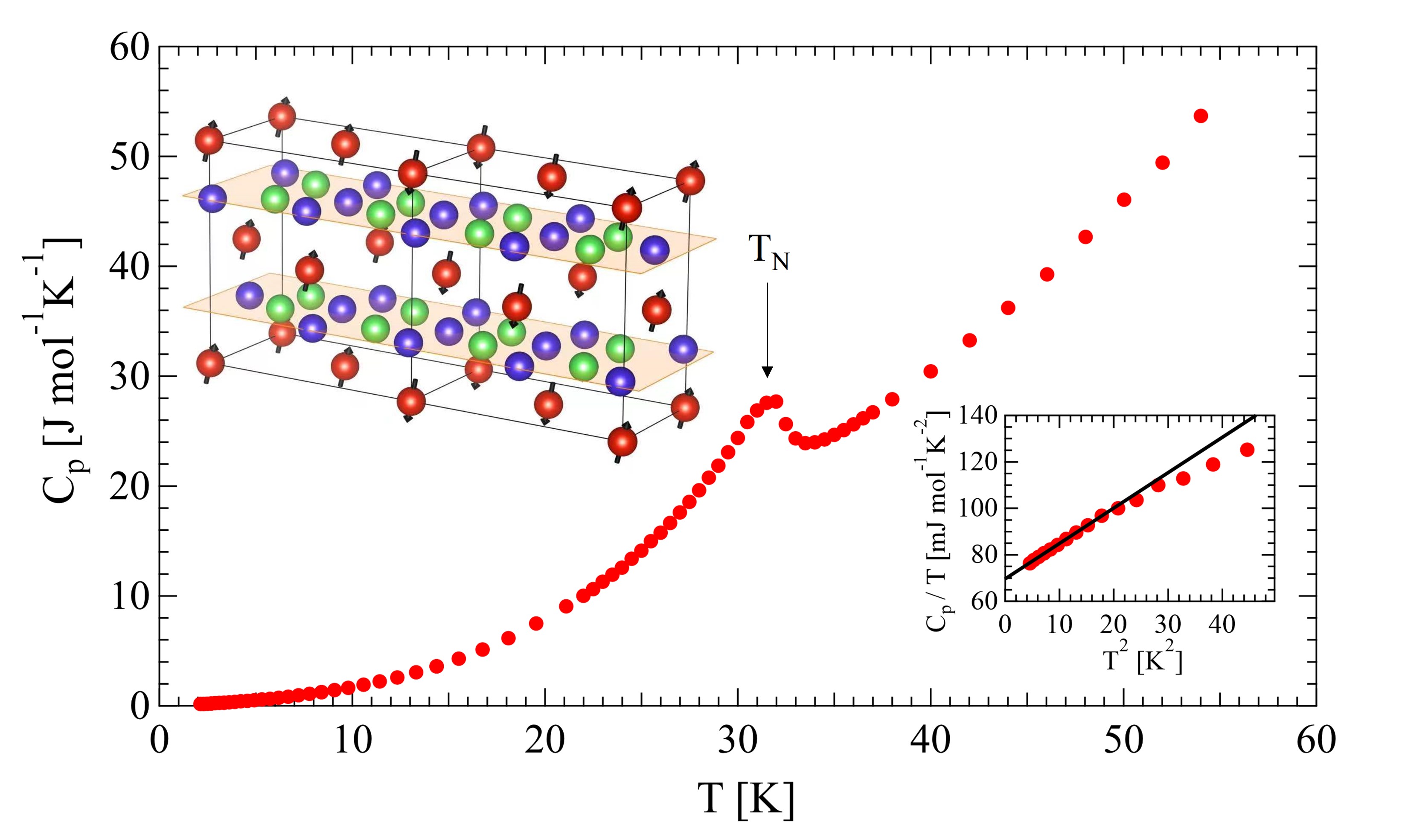}
	\caption{Temperature dependence of the isobaric specific heat, $C_P$ of a \VS{} powder sample. Top inset: the V ion lattice of \VS{}, adapted from Ref. \cite{Kaw75}. Blue, green and red spheres indicate V2 and V3 sites in the 1T-CdI$_2$ type VS$_2$ layers, highlighted in orange, and the V1 site intercalated in between these planes. Arrows represent the AFM ordered moments of the V1 ions. Bottom inset: the $C_P/T$ vs. $T^2$ data at low temperatures. The black line is a linear fit (see Eq. \ref{CpT}).}
	\label{Cp_structure}
\end{figure}

In order to elucidate the above controversy, here we consider the intercalated layered TM dichalcogenide V$_{1+x}$S$_2$ as a model system of Kondo lattice where the above alternative HF mechanisms are absent. Our choice is dictated by a number of favorable conditions. (i) Owing to a large separation between intercalated V ions, the $d$-states of these ions are expected to be localized, which is supported by photoemission spectroscopy \cite{fuj91} and \textit{ab initio} calculations \cite{suz93,kne98,koo01,niu20}, and thus to behave as Kondo scatterers for the conduction electrons in the VS$_2$ layers; (ii) different from other layered dichalcogenides, in V$_{1+x}$S$_2$, the concentration of intercalated ions can vary up to $x=0.5$, thus enabling to tune the RKKY and Kondo interactions. We focus on the $x=0.25$ composition, or \VS{} \cite{Kaw75}, where the intercalated ions in the V1 site form a hexagonal sublattice, a favorable condition to create a Kondo lattice (see Figure \ref{Cp_structure}). A picture of dominant RKKY interaction for \VS{} is supported by the observation of a long-range AFM order of the V1 ions at $T_N = 32$ K \cite{fun81} and by a theoretical study on intercalated TiS$_2$ \cite{suz93}. (iii) Consistent with the Doniach phase diagram, upon increasing the concentration of intercalated V ions up to $x=0.5$ and partially substituting them for nonmagnetic Cu ions, the AFM order is suppressed concomitant to the appearance of a Kondo behavior \cite{gau19}, also reported in the related system VSe$_2$ \cite{bar17}. The above results indicate that the RKKY and Kondo interactions are both present in V$_{1+x}$S$_2$, which prompts us to explore their competition by using pressure as control parameter, as done in the aforementioned studies on $f$-electron systems, with the expectation to suppress the AFM order at a critical value of pressure, $P_c$. 

The \VS{} samples studied in this work are single-phase powders and single crystals grown by the chemical vapor transport method, as described previously \cite{sae74,Kaw75}. We reproducibly obtained high-quality single crystals with either platelet- or needle-like shapes depending on growth conditions, i.e. by applying a temperature gradient of 750-900 $^\circ$C or 950-1010 $^\circ$C, respectively. After having confirmed the high crystalline quality by means of x-ray diffraction, we investigated systematically the samples by means of isobaric specific heat $C_P$ and magnetic susceptibility, $\chi$, measurements as a function of temperature, $T$, in the 2-300 K range and by electrical resistivity, $\varrho$, measurements as a function of temperature in the same range at ambient and high pressures up to 10 GPa.

Figure \ref{Cp_structure} shows the $C_P(T)$ curve measured on a \VS{} powder sample using a $2 \tau$ relaxation technique in a Quantum Design Physical Properties Measurement System (PPMS). In agreement with recent data \cite{niu20}, note a jump at the AFM transition, $T_N = 32$ K, consistent with previous magnetic susceptibility \cite{Kaw75,nis77} and neutron diffraction \cite{fun81} studies. Below $T_N$, the curve follows the conventional behavior:

\begin{equation}\label{CpT} 
	C_P/T = \gamma + \beta T^2
\end{equation}

where the quadratic term represents the lattice contribution and the electronic term, $\gamma = \frac{\pi^2}{3}k_B^2 D$, is directly proportional to the effective mass, $m^{\ast}$, and to the Fermi wave vector, $k_F$, i.e. $D=m^{\ast}\frac{k_F}{\pi^2\hbar^2}$. Our analysis indicates that an additional spin-wave contribution proposed recently \cite{niu20} is not necessary to explain the data. We obtain $\gamma \sim 70$ mJ K$^{-2}$ per mole of \VS{} formula unit or $\sim 14$ mJ K$^{-2}$ per mole of V, an unusually high value for a TM sulfide, which indicates a sizable enhancement of the effective mass, especially considering the small Fermi surface typical of a semimetal. To the best of our knowledge, the only TM sulfides displaying similar values are \CVS{} \cite{gau19}, obtained by partially substituting the intercalated V ion for Cu in the present V$_{1+x}$S$_2$ system for $x=0.5$, and another layered dichalcogenide, $M_x$TiS$_2$, intercalated with various TM ions, $M$ \cite{ino86}. We recall that comparably large $\gamma$ values are typical of correlated TM oxides, such as Sr$_2$RuO$_4$ \cite{mae97}, LiV$_2$O$_4$ \cite{joh99} and Ca$_3$Co$_4$O$_9$ \cite{lim05}. A further indication of mass enhancement in \VS{} is that, from the above relation, we obtain $D = 30$ states eV$^{-1}$ (f.u.)$^{-1}$ a value 3-5 times larger than that predicted by \textit{ab initio} calculations based on density functional theory (DFT) \cite{kne98,koo01,niu20}.

\begin{figure}[ht]
	\centering
	\includegraphics[width =\columnwidth]{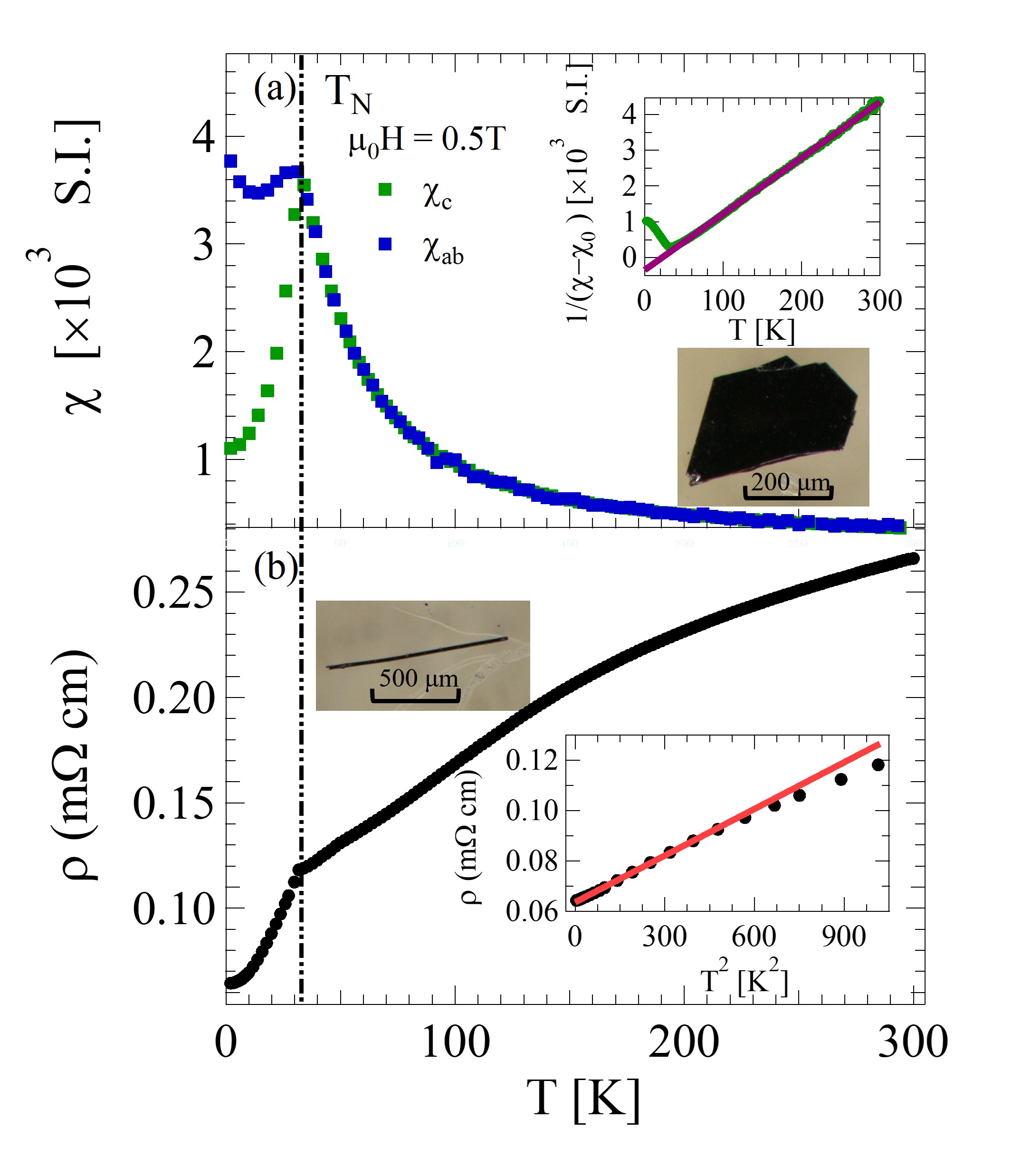}
	\caption{a: zero-field cooling (ZFC) magnetic susceptibility of a plate-like \VS{} single crystal measured in a 0.5 T field parallel (blue) and perpendicular (green) to the VS$_2$ $ab$-planes. The data are corrected for the core diamagnetic susceptibility (see text). The field-cooling curves do not differ from the ZFC ones. Inset: Curie-Weiss behavior above $T_N$. b: out-of-plane ($c$-axis) resistivity curve of a needle-like crystal. The inset shows the quadratic dependence below $T_N$. The pictures in (a) and (b) show the single crystals measured.}
	\label{V5S8_chi_rho}
\end{figure}

Our magnetic susceptibility results provide further evidence of a sizable mass enhancement. Indeed, the Pauli susceptibility is also proportional to $D$ and $m^{\ast}$, i.e. $\chi_P=\mu_0 \mu_B^2 D$, where $\mu_B$ is the Bohr magneton. In Figure \ref{V5S8_chi_rho}, we plot the $\chi(T)$ curves taken on a representative platelet-like \VS{} single crystal in the 2-300 K range using a Quantum Design Magnetic Properties Measurement System (MPMS) equipped with a Reciprocating Sample Option (RSO). The data are shown after correction for the core susceptibility, $\chi_{core} \approx -1.4 \times 10^{-5}$ S.I. \cite{hel76}. Above $T_N$, we find an isotropic Curie-Weiss behavior, in agreement with a picture of local moments in the intercalated V1 site. Below $T_N$, the anisotropy of the response is typical of an AFM structure with moments predominantly aligned along the out-of-plane $c$-direction, in agreement with the structural model proposed in an earlier neutron diffraction study \cite{fun81}. We then analyze the data using the expression:

\begin{equation}
   \chi(T) = \chi_0 + \frac{C}{T-\Theta}
\end{equation}

where $\chi_0$ is the sum of the Pauli paramagnetism, $\chi_P$, and Landau diamagnetism, $\chi_L = -\frac{1}{3}\frac{m}{m^\ast}\chi_P$, and the second term is the Curie-Weiss term. We obtain $C$ = 6.3(2) $\times 10^{-2}$ K, $\chi_0 =$ 1.40(3) $\times 10^{-4}$ S.I. (or 1.42 $\times 10^{-3}$ cm$^3$ mol$^{-1}$) and $\Theta = 23$ K. The value of $C$ is comparable to previously published values ranging from 5.82 $\times 10^{-2}$ K to 7.91 $\times 10^{-2}$ K \cite{noz78,nis77}. Assuming that only the V1 ions carry a local moment, we obtain an effective moment, $\mu =$ 2.2 $\mu_B$ per V1, in excellent agreement with neutron diffraction \cite{fun81} and magnetization studies \cite{nak00}. This value falls in between the values 1.73 $\mu_B$ and 2.82 $\mu_B$ expected for the V$^{4+}$ and V$^{3+}$ ions in the $3d^1$ ($S = 1/2$) and $3d^2$ ($S = 1$) electronic configurations, respectively, consistent with the nominal average valence V$^{3.2+}$ and with photoemission spectroscopy \cite{fuj91}.

By neglecting $\chi_L$ owing to the enhanced effective mass, we obtain $\chi_P \approx \chi_0 \approx 1.5 \times 10^{-3}$ cm$^3$ mol$^{-1}$. As in the case of $\gamma$, also this value is unusually large for a TM sulfide and rather typical of the aforementioned correlated TM oxides. A picture of correlated metal for \VS{} is confirmed by the universal prediction for the Sommerfeld–Wilson ratio, $R_W = \frac{4}{3\mu_0}\left(\frac{\pi k_B}{g \mu_B}\right)^2\frac{\chi_P}{\gamma}$, where $g$ is the gyromagnetic ratio of the electron. In a free-electron gas, both $\chi_P$ and $\gamma$ are proportional to $D$, so $R_W$ = 1, while many-body effects renormalize this value up to 2 \cite{lee86}, consistent with the result of renormalization group theory \cite{wil75,noz80}. For \VS{}, we obtain $R_W = 1.5$, which confirms that many-body effects are important.

A scenario of significant mass enhancement is further supported by the analysis of the $\varrho$ data. Figure \ref{V5S8_chi_rho} shows the low-temperature dependence of the out-of-plane $c$-axis resistivity, $\varrho_c(T)$, of a representative single crystal of dimensions $1000 \times 60 \times 60 \mu$m$^3$ at ambient pressure. The needle-like shape of the sample enabled us to conveniently measure $\varrho_c$ using a four-point method in the bar configuration. Note a clear quadratic dependence that extends up to the AFM transition at $T_N=32$ K, where the curve displays a sharp kink, in agreement with previous reports \cite{noz75,niu20}. A data fit within the 2-25 K range of the type $\varrho_c(T) = \varrho_0 + AT^2$ yields a quadratic resistivity coefficient $A = 6 \times 10^{-2} \mu\Omega$ cm K$^{-2}$, three times larger than the in-plane value \cite{niu20}, which suggests a modest anisotropy of the transport properties. As well established for typical HFs, such as CeCu$_2$Ge$_2$ \cite{kne97}, Ce$_4$Pt$_{12}$Sn$_{25}$ \cite{lee10} and Ce$_2$Ni$_3$Ge$_5$ \cite{hos00}, such a quadratic dependence is the signature of electron-electron scattering in a coherent FL, so $A \sim D^2$ is a further measure of $m^{\ast}$. Consistent with the previous analysis of $\gamma$ and $\chi_P$, the $A$ value of \VS{} is unusually large for a TM sulfide and comparable with that of correlated TM oxides, such as Sr$_2$RuO$_4$ \cite{mae97} and V$_2$O$_3$ \cite{mcw69}. We confirm that the quadratic term arises from electronic correlations by calculating the Kadowaki-Woods ratio, $R_{KW} = A/\gamma^2$. We obtain $R_{KW}=1.2 \times 10^{-5} \mu\Omega$ cm(mol K mJ$^{-1}$)$^{2}$, which follows well the trend of typical HFs \cite{kad86} in the intermediate mass renormalization region of correlated TM oxides.

The room temperature value, 0.2 m$\Omega$ cm, and the temperature dependence of $\varrho_{c}$ above $T_N$ are similar to those of the in-plane resistivity, $\varrho_{ab}$ \cite{niu20}, which confirms a picture of 3D transport properties with a $\sim 3$ anisotropy factor. Notable is an indication of saturation above 150 K, as expected in the Ioffe-Regel limit $k_F\ell \sim 1$ reached in bad metals with short mean-free paths, $\ell$, or small carrier densities (i.e. small $k_F$). This interpretation is consistent with the comparatively high resistivity value and the modest residual resistivity ratio, $RRR= \varrho(300 {\rm K})/\varrho(2 {\rm K}) = 4.4$. Nozaki \textit{et al.} proposed an interpretation of the hump at 150 K in terms of thermally activated n-type carriers within a two-band model \cite{noz75}. 

In summary, all the above results consistently point at a scenario of an itinerant AFM phase, where the electrons in the metallic VS$_2$ layers are correlated. We account for this scenario within the frame of a generalized Kondo lattice where the RKKY interaction not only induces the AFM order of the local moments of the intercalated ions, but also renormalizes the conduction band. This picture is consistent with the fact that layered dichalcogenides with low concentrations of intercalated ions (i.e. with no Kondo lattice), do not exhibit either magnetic order or mass enhancement effects.

Our point is that the HF properties of such AFM phase are unusual within the frame of the Doniach phase diagram of traditional $f$-electron HFs, where HF properties are found only in the paramagnetic phase, while the conduction electrons in the magnetic phase are quasi-free like \cite{fri09}. We explain this difference by the narrow conduction $d$-band of \VS{}, which is prone to exhibit correlation effects, different from the broad conduction $s$- or $p$-bands of $f$-electron HFs. For \VS{}, we should then generalize the Doniach phase diagram as shown in Figure \ref{phase-diagram}, by including the heavy FL (HFL) character in the metallic AFM phase in the low-$h$ side of the diagram.  

The question is whether the QCP scenario describing the competition between RKKY and Kondo interactions in $f$-electron systems is applicable to \VS{}. In the positive, we expect a suppression of the AFM order at a critical pressure, $P_c$. To verify this scenario, we measured the $c$-axis resistivity of the same \VS{} single crystal of Figure 2b in a Bridgman cell up to 10 GPa \cite{Nic15}. We determined the $P$-value from the superconducting critical temperature, $T_c$, of a high-purity Pb sample placed near the \VS{} sample and using the $P$-dependence of $T_c$ as pressure calibration.

\begin{figure}
	\centering 
	\includegraphics[width = \columnwidth]{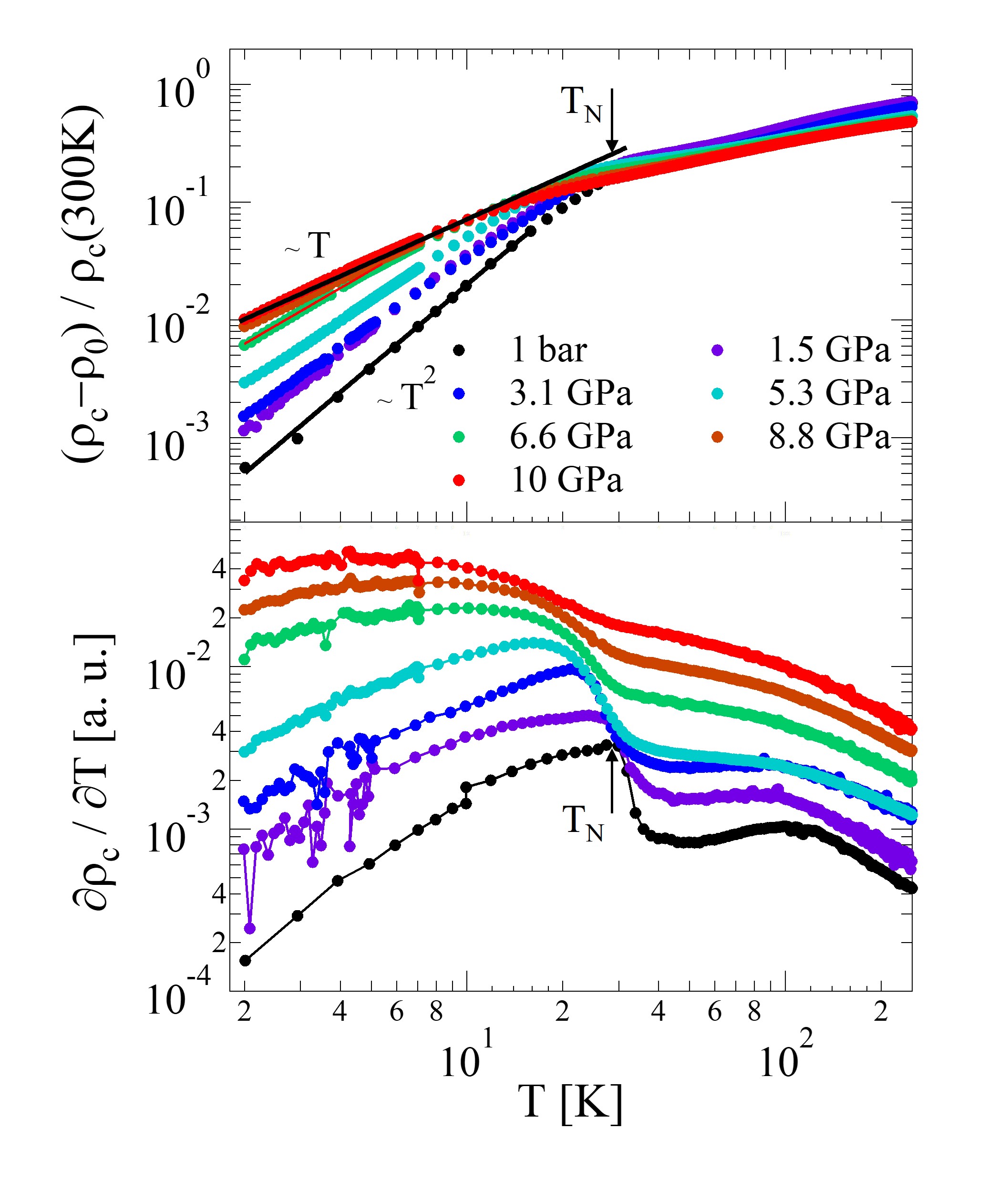}
	\caption{(a) Evolution of the normalized $c$-axis resistivity, $\varrho_c(T)$, of a representative \VS{} single crystal with pressure. Black lines are a fit to the power law $(\varrho_c(T)-\varrho_{0})/\varrho_c$(300K) $= aT^n$, where $\varrho_0$ is the residual resistivity and $a$ and $n$ are free parameters. The AFM ordering temperatures, $T_N$, are determined from the jump in the derivative curves of panel (b) that are shifted vertically for clarity.}  
	\label{V5S8_pressure}
\end{figure}

\begin{figure}[ht]
	\centering 
	\includegraphics[width = \columnwidth]{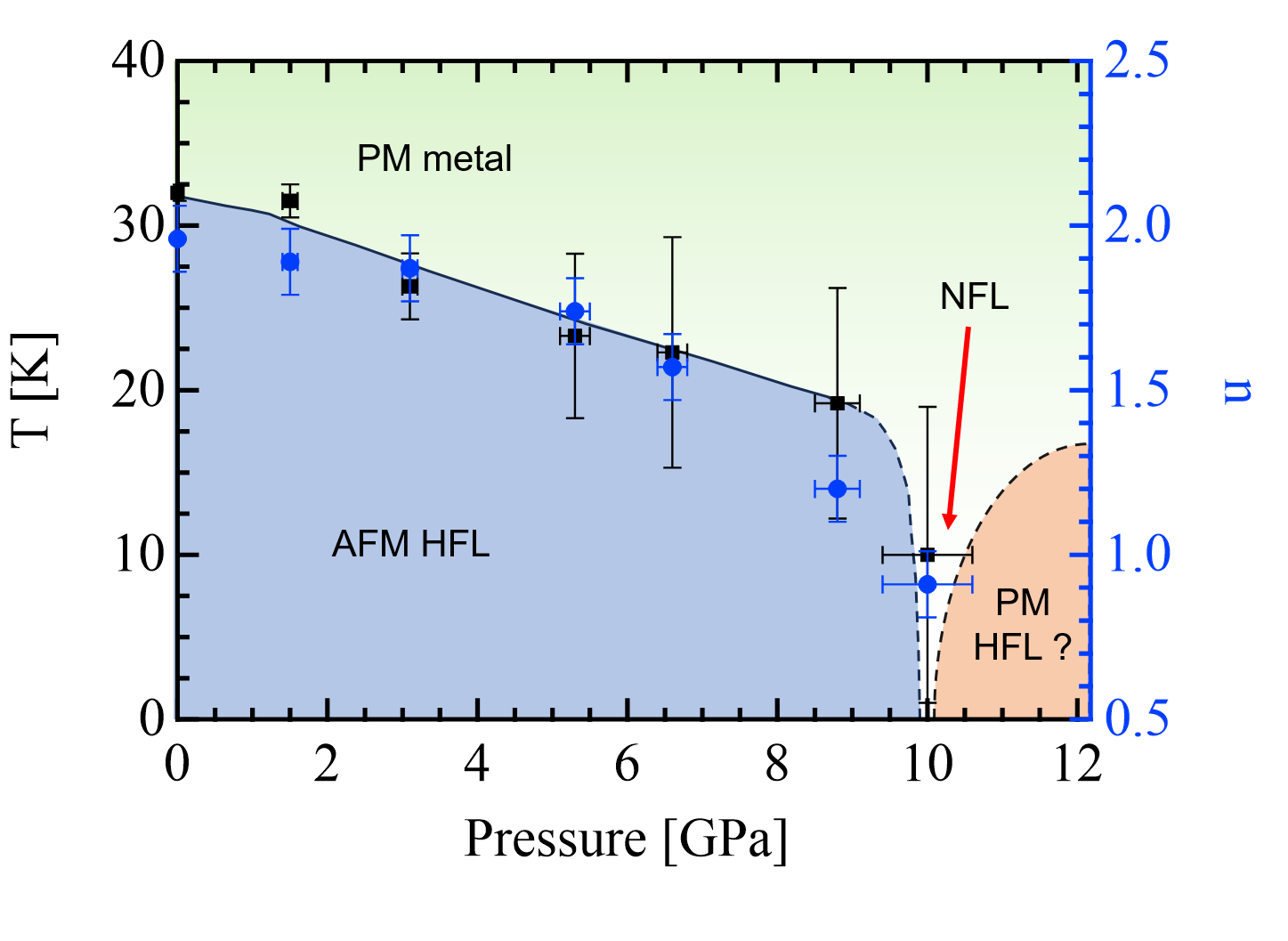}
	\caption{Generalized Doniach phase diagram proposed for \VS. Black squares (left axis) indicate $T_N$, as determined from the jump in the derivative curves, $\frac{\partial \varrho_c}{\partial T}$, of Figure \ref{V5S8_pressure}. Blue circles (right axis) indicate the exponent $n$ of the power law, $\varrho_c(T) = \varrho_{0} + aT^n$.}  
	\label{phase-diagram}
\end{figure}

The $\varrho_c(T)$ data in Figure \ref{V5S8_pressure} show a progressive suppression of the AFM order with pressure, as better seen in the derivative curves, $\frac{\partial\varrho_c}{\partial T}$, where the AFM transition appears as a jump. Note that the jump remains well defined up to 3.5 GPa and then broadens, until it progressively disappears at $P_c \approx$ 10 GPa. In Figure \ref{phase-diagram}, note a gradual decrease of $T_N$ until 9 GPa, followed by a rapid drop at 10 GPa. Remarkably, this behavior is mirrored by the behavior of the power law of $\varrho_c(T)$. By analyzing the data using the power law $\varrho_c(T) = \varrho_0 + AT^n$, we find that the exponent $n$ smoothly decreases from 2 at ambient pressure to 1.5 at 6.6 GPa and then quickly drops to 1, again at $P_c \approx$ 10 GPa, signature of a breakdown of the FL phase (see Figure \ref{phase-diagram}).  

Interestingly, a similar pressure-induced suppression of magnetic order accompanied by the appearance of a linear behavior of the resistivity has been reported in the $f$-electron HF compound CeRu$_2$Ge$_2$ \cite{sue99}, which has been interpreted as the onset of critical fluctuations near the QCP and thus as a signature of non-FL (NFL) behavior. This similarity supports a scenario of QCP for \VS{} as well. The important difference is that, as mentioned before, in $f$-electron systems, the quadratic dependence of the resistivity signature of HF regime is observed in the high-$h$ side of the Doniach phase diagram.

In conclusion, our finding of QCP in \VS{} confirms a scenario of Kondo lattice produced by the local moments of the intercalated V ions, which elucidates the long-standing controversy regarding the suitability of $d$-electron systems to host an effective Kondo interaction. The observation of an unusual coexistence of heavy quasiparticles in the metallic VS$_2$ layers and of a AFM order of the intercalated ions suggests a generalized Doniach phase diagram where, different from the case of traditional $f$-electron systems, the scattering by the local moments induces a significant renormalization of the conduction band in the magnetic phase, i.e. on the low-$h$ side of the QCP. The question arises whether, in the paramagnetic phase at higher pressures, i.e. in the strong coupling regime on the high-$h$ side of the QCP, the heavy quasiparticles survive, as in the above $f$-electron systems, or form a NFL phase.

\begin{acknowledgments}
The authors thank Hidenori Takagi for useful discussions and gratefully acknowledge the financial support provided by the Chinese Scholarship Council (Grant No. 201608070037) and by the Laboratory of Excellence MATISSE within the frame of the "Investissements d'Avenir Programme" of the French Ministry of University and Research (MESRI) under reference ANR-11-IDEX-0004-02.
\end{acknowledgments}

\bibliographystyle{apsrev4-1}
\bibliography{V5S8_biblio}

\begin{thebibliography}{50}%
\makeatletter
\providecommand \@ifxundefined [1]{%
 \@ifx{#1\undefined}
}%
\providecommand \@ifnum [1]{%
 \ifnum #1\expandafter \@firstoftwo
 \else \expandafter \@secondoftwo
 \fi
}%
\providecommand \@ifx [1]{%
 \ifx #1\expandafter \@firstoftwo
 \else \expandafter \@secondoftwo
 \fi
}%
\providecommand \natexlab [1]{#1}%
\providecommand \enquote  [1]{``#1''}%
\providecommand \bibnamefont  [1]{#1}%
\providecommand \bibfnamefont [1]{#1}%
\providecommand \citenamefont [1]{#1}%
\providecommand \href@noop [0]{\@secondoftwo}%
\providecommand \href [0]{\begingroup \@sanitize@url \@href}%
\providecommand \@href[1]{\@@startlink{#1}\@@href}%
\providecommand \@@href[1]{\endgroup#1\@@endlink}%
\providecommand \@sanitize@url [0]{\catcode `\\12\catcode `\$12\catcode
  `\&12\catcode `\#12\catcode `\^12\catcode `\_12\catcode `\%12\relax}%
\providecommand \@@startlink[1]{}%
\providecommand \@@endlink[0]{}%
\providecommand \url  [0]{\begingroup\@sanitize@url \@url }%
\providecommand \@url [1]{\endgroup\@href {#1}{\urlprefix }}%
\providecommand \urlprefix  [0]{URL }%
\providecommand \Eprint [0]{\href }%
\providecommand \doibase [0]{https://doi.org/}%
\providecommand \selectlanguage [0]{\@gobble}%
\providecommand \bibinfo  [0]{\@secondoftwo}%
\providecommand \bibfield  [0]{\@secondoftwo}%
\providecommand \translation [1]{[#1]}%
\providecommand \BibitemOpen [0]{}%
\providecommand \bibitemStop [0]{}%
\providecommand \bibitemNoStop [0]{.\EOS\space}%
\providecommand \EOS [0]{\spacefactor3000\relax}%
\providecommand \BibitemShut  [1]{\csname bibitem#1\endcsname}%
\let\auto@bib@innerbib\@empty
\bibitem [{\citenamefont {Kondo}(1964)}]{kon64}%
  \BibitemOpen
  \bibfield  {author} {\bibinfo {author} {\bibfnamefont {J.}~\bibnamefont
  {Kondo}},\ }\href {https://doi.org/10.1143/PTP.32.37} {\bibfield  {journal}
  {\bibinfo  {journal} {Progress of Theoretical Physics}\ }\textbf {\bibinfo
  {volume} {32}},\ \bibinfo {pages} {37} (\bibinfo {year} {1964})}\BibitemShut
  {NoStop}%
\bibitem [{\citenamefont {Stewart}(1984)}]{ste84}%
  \BibitemOpen
  \bibfield  {author} {\bibinfo {author} {\bibfnamefont {G.~R.}\ \bibnamefont
  {Stewart}},\ }\href {https://doi.org/10.1103/RevModPhys.56.755} {\bibfield
  {journal} {\bibinfo  {journal} {Rev. Mod. Phys.}\ }\textbf {\bibinfo {volume}
  {56}},\ \bibinfo {pages} {755} (\bibinfo {year} {1984})}\BibitemShut
  {NoStop}%
\bibitem [{\citenamefont {Andres}\ \emph {et~al.}(1975)\citenamefont {Andres},
  \citenamefont {Graebner},\ and\ \citenamefont {Ott}}]{and75}%
  \BibitemOpen
  \bibfield  {author} {\bibinfo {author} {\bibfnamefont {K.}~\bibnamefont
  {Andres}}, \bibinfo {author} {\bibfnamefont {J.~E.}\ \bibnamefont
  {Graebner}},\ and\ \bibinfo {author} {\bibfnamefont {H.~R.}\ \bibnamefont
  {Ott}},\ }\href {https://doi.org/10.1103/PhysRevLett.35.1779} {\bibfield
  {journal} {\bibinfo  {journal} {Phys. Rev. Lett.}\ }\textbf {\bibinfo
  {volume} {35}},\ \bibinfo {pages} {1779} (\bibinfo {year}
  {1975})}\BibitemShut {NoStop}%
\bibitem [{\citenamefont {Scheffler}\ \emph {et~al.}(2005)\citenamefont
  {Scheffler}, \citenamefont {Dressel}, \citenamefont {Jourdan},\ and\
  \citenamefont {Adrian}}]{sch05}%
  \BibitemOpen
  \bibfield  {author} {\bibinfo {author} {\bibfnamefont {M.}~\bibnamefont
  {Scheffler}}, \bibinfo {author} {\bibfnamefont {M.}~\bibnamefont {Dressel}},
  \bibinfo {author} {\bibfnamefont {M.}~\bibnamefont {Jourdan}},\ and\ \bibinfo
  {author} {\bibfnamefont {H.}~\bibnamefont {Adrian}},\ }\href
  {https://doi.org/10.1038/nature04232} {\bibfield  {journal} {\bibinfo
  {journal} {Nature}\ }\textbf {\bibinfo {volume} {438}},\ \bibinfo {pages}
  {1135} (\bibinfo {year} {2005})}\BibitemShut {NoStop}%
\bibitem [{\citenamefont {Doniach}(1977)}]{don77}%
  \BibitemOpen
  \bibfield  {author} {\bibinfo {author} {\bibfnamefont {S.}~\bibnamefont
  {Doniach}},\ }\bibinfo {title} {Phase diagram for the kondo lattice},\ in\
  \href {https://doi.org/10.1007/978-1-4615-8816-0_15} {\emph {\bibinfo
  {booktitle} {Valence Instabilities and Related Narrow-Band Phenomena}}},\
  \bibinfo {editor} {edited by\ \bibinfo {editor} {\bibfnamefont {R.~D.}\
  \bibnamefont {Parks}}}\ (\bibinfo  {publisher} {Springer US},\ \bibinfo
  {address} {Boston, MA},\ \bibinfo {year} {1977})\ pp.\ \bibinfo {pages}
  {169--176}\BibitemShut {NoStop}%
\bibitem [{\citenamefont {S\"ullow}\ \emph {et~al.}(1999)\citenamefont
  {S\"ullow}, \citenamefont {Aronson}, \citenamefont {Rainford},\ and\
  \citenamefont {Haen}}]{sue99}%
  \BibitemOpen
  \bibfield  {author} {\bibinfo {author} {\bibfnamefont {S.}~\bibnamefont
  {S\"ullow}}, \bibinfo {author} {\bibfnamefont {M.~C.}\ \bibnamefont
  {Aronson}}, \bibinfo {author} {\bibfnamefont {B.~D.}\ \bibnamefont
  {Rainford}},\ and\ \bibinfo {author} {\bibfnamefont {P.}~\bibnamefont
  {Haen}},\ }\href {https://doi.org/10.1103/PhysRevLett.82.2963} {\bibfield
  {journal} {\bibinfo  {journal} {Phys. Rev. Lett.}\ }\textbf {\bibinfo
  {volume} {82}},\ \bibinfo {pages} {2963} (\bibinfo {year}
  {1999})}\BibitemShut {NoStop}%
\bibitem [{\citenamefont {L\"ohneysen}\ \emph {et~al.}(1994)\citenamefont
  {L\"ohneysen}, \citenamefont {Pietrus}, \citenamefont {Portisch},
  \citenamefont {Schlager}, \citenamefont {Schr\"oder}, \citenamefont {Sieck},\
  and\ \citenamefont {Trappmann}}]{loe94}%
  \BibitemOpen
  \bibfield  {author} {\bibinfo {author} {\bibfnamefont {H.~v.}\ \bibnamefont
  {L\"ohneysen}}, \bibinfo {author} {\bibfnamefont {T.}~\bibnamefont
  {Pietrus}}, \bibinfo {author} {\bibfnamefont {G.}~\bibnamefont {Portisch}},
  \bibinfo {author} {\bibfnamefont {H.~G.}\ \bibnamefont {Schlager}}, \bibinfo
  {author} {\bibfnamefont {A.}~\bibnamefont {Schr\"oder}}, \bibinfo {author}
  {\bibfnamefont {M.}~\bibnamefont {Sieck}},\ and\ \bibinfo {author}
  {\bibfnamefont {T.}~\bibnamefont {Trappmann}},\ }\href
  {https://doi.org/10.1103/PhysRevLett.72.3262} {\bibfield  {journal} {\bibinfo
   {journal} {Phys. Rev. Lett.}\ }\textbf {\bibinfo {volume} {72}},\ \bibinfo
  {pages} {3262} (\bibinfo {year} {1994})}\BibitemShut {NoStop}%
\bibitem [{\citenamefont {Friedemann}\ \emph {et~al.}(2009)\citenamefont
  {Friedemann}, \citenamefont {Westerkamp}, \citenamefont {Brando},
  \citenamefont {Oeschler}, \citenamefont {Wirth}, \citenamefont {Gegenwart},
  \citenamefont {Krellner}, \citenamefont {Geibel},\ and\ \citenamefont
  {Steglich}}]{fri09}%
  \BibitemOpen
  \bibfield  {author} {\bibinfo {author} {\bibfnamefont {S.}~\bibnamefont
  {Friedemann}}, \bibinfo {author} {\bibfnamefont {T.}~\bibnamefont
  {Westerkamp}}, \bibinfo {author} {\bibfnamefont {M.}~\bibnamefont {Brando}},
  \bibinfo {author} {\bibfnamefont {N.}~\bibnamefont {Oeschler}}, \bibinfo
  {author} {\bibfnamefont {S.}~\bibnamefont {Wirth}}, \bibinfo {author}
  {\bibfnamefont {P.}~\bibnamefont {Gegenwart}}, \bibinfo {author}
  {\bibfnamefont {C.}~\bibnamefont {Krellner}}, \bibinfo {author}
  {\bibfnamefont {C.}~\bibnamefont {Geibel}},\ and\ \bibinfo {author}
  {\bibfnamefont {F.}~\bibnamefont {Steglich}},\ }\href
  {https://doi.org/10.1038/nphys1299} {\bibfield  {journal} {\bibinfo
  {journal} {Nature Physics}\ }\textbf {\bibinfo {volume} {5}},\ \bibinfo
  {pages} {465} (\bibinfo {year} {2009})}\BibitemShut {NoStop}%
\bibitem [{\citenamefont {Trovarelli}\ \emph {et~al.}(2000)\citenamefont
  {Trovarelli}, \citenamefont {Geibel}, \citenamefont {Mederle}, \citenamefont
  {Langhammer}, \citenamefont {Grosche}, \citenamefont {Gegenwart},
  \citenamefont {Lang}, \citenamefont {Sparn},\ and\ \citenamefont
  {Steglich}}]{tro00}%
  \BibitemOpen
  \bibfield  {author} {\bibinfo {author} {\bibfnamefont {O.}~\bibnamefont
  {Trovarelli}}, \bibinfo {author} {\bibfnamefont {C.}~\bibnamefont {Geibel}},
  \bibinfo {author} {\bibfnamefont {S.}~\bibnamefont {Mederle}}, \bibinfo
  {author} {\bibfnamefont {C.}~\bibnamefont {Langhammer}}, \bibinfo {author}
  {\bibfnamefont {F.~M.}\ \bibnamefont {Grosche}}, \bibinfo {author}
  {\bibfnamefont {P.}~\bibnamefont {Gegenwart}}, \bibinfo {author}
  {\bibfnamefont {M.}~\bibnamefont {Lang}}, \bibinfo {author} {\bibfnamefont
  {G.}~\bibnamefont {Sparn}},\ and\ \bibinfo {author} {\bibfnamefont
  {F.}~\bibnamefont {Steglich}},\ }\href
  {https://doi.org/10.1103/PhysRevLett.85.626} {\bibfield  {journal} {\bibinfo
  {journal} {Phys. Rev. Lett.}\ }\textbf {\bibinfo {volume} {85}},\ \bibinfo
  {pages} {626} (\bibinfo {year} {2000})}\BibitemShut {NoStop}%
\bibitem [{\citenamefont {Stewart}(2001)}]{ste01}%
  \BibitemOpen
  \bibfield  {author} {\bibinfo {author} {\bibfnamefont {G.~R.}\ \bibnamefont
  {Stewart}},\ }\href {https://doi.org/10.1103/RevModPhys.73.797} {\bibfield
  {journal} {\bibinfo  {journal} {Rev. Mod. Phys.}\ }\textbf {\bibinfo {volume}
  {73}},\ \bibinfo {pages} {797} (\bibinfo {year} {2001})}\BibitemShut
  {NoStop}%
\bibitem [{\citenamefont {Coleman}\ and\ \citenamefont
  {Schofield}(2005)}]{col05}%
  \BibitemOpen
  \bibfield  {author} {\bibinfo {author} {\bibfnamefont {P.}~\bibnamefont
  {Coleman}}\ and\ \bibinfo {author} {\bibfnamefont {A.~J.}\ \bibnamefont
  {Schofield}},\ }\href {https://doi.org/10.1038/nature03279} {\bibfield
  {journal} {\bibinfo  {journal} {Nature}\ }\textbf {\bibinfo {volume} {433}},\
  \bibinfo {pages} {226} (\bibinfo {year} {2005})}\BibitemShut {NoStop}%
\bibitem [{\citenamefont {L\"ohneysen}\ \emph {et~al.}(2007)\citenamefont
  {L\"ohneysen}, \citenamefont {Rosch}, \citenamefont {Vojta},\ and\
  \citenamefont {W\"olfle}}]{loe07}%
  \BibitemOpen
  \bibfield  {author} {\bibinfo {author} {\bibfnamefont {H.~v.}\ \bibnamefont
  {L\"ohneysen}}, \bibinfo {author} {\bibfnamefont {A.}~\bibnamefont {Rosch}},
  \bibinfo {author} {\bibfnamefont {M.}~\bibnamefont {Vojta}},\ and\ \bibinfo
  {author} {\bibfnamefont {P.}~\bibnamefont {W\"olfle}},\ }\href
  {https://doi.org/10.1103/RevModPhys.79.1015} {\bibfield  {journal} {\bibinfo
  {journal} {Rev. Mod. Phys.}\ }\textbf {\bibinfo {volume} {79}},\ \bibinfo
  {pages} {1015} (\bibinfo {year} {2007})}\BibitemShut {NoStop}%
\bibitem [{\citenamefont {Gegenwart}\ \emph {et~al.}(2008)\citenamefont
  {Gegenwart}, \citenamefont {Si},\ and\ \citenamefont {Steglich}}]{geg08}%
  \BibitemOpen
  \bibfield  {author} {\bibinfo {author} {\bibfnamefont {P.}~\bibnamefont
  {Gegenwart}}, \bibinfo {author} {\bibfnamefont {Q.}~\bibnamefont {Si}},\ and\
  \bibinfo {author} {\bibfnamefont {F.}~\bibnamefont {Steglich}},\ }\href
  {https://doi.org/10.1038/nphys892} {\bibfield  {journal} {\bibinfo  {journal}
  {Nature Physics}\ }\textbf {\bibinfo {volume} {4}},\ \bibinfo {pages} {186}
  (\bibinfo {year} {2008})}\BibitemShut {NoStop}%
\bibitem [{\citenamefont {Krug~von Nidda}\ \emph {et~al.}(2003)\citenamefont
  {Krug~von Nidda}, \citenamefont {Bulla}, \citenamefont {B{\"u}ttgen},
  \citenamefont {Heinrich},\ and\ \citenamefont {Loidl}}]{kru03}%
  \BibitemOpen
  \bibfield  {author} {\bibinfo {author} {\bibfnamefont {H.~A.}\ \bibnamefont
  {Krug~von Nidda}}, \bibinfo {author} {\bibfnamefont {R.}~\bibnamefont
  {Bulla}}, \bibinfo {author} {\bibfnamefont {N.}~\bibnamefont {B{\"u}ttgen}},
  \bibinfo {author} {\bibfnamefont {M.}~\bibnamefont {Heinrich}},\ and\
  \bibinfo {author} {\bibfnamefont {A.}~\bibnamefont {Loidl}},\ }\href
  {https://doi.org/10.1140/epjb/e2003-00237-9} {\bibfield  {journal} {\bibinfo
  {journal} {The European Physical Journal B - Condensed Matter and Complex
  Systems}\ }\textbf {\bibinfo {volume} {34}},\ \bibinfo {pages} {399}
  (\bibinfo {year} {2003})}\BibitemShut {NoStop}%
\bibitem [{\citenamefont {Niemann}\ \emph {et~al.}(2023)\citenamefont
  {Niemann}, \citenamefont {Wu}, \citenamefont {Oka}, \citenamefont {Hirai},
  \citenamefont {Wang}, \citenamefont {Suyolcu}, \citenamefont {Kim},
  \citenamefont {van Aken},\ and\ \citenamefont {Takagi}}]{nie23}%
  \BibitemOpen
  \bibfield  {author} {\bibinfo {author} {\bibfnamefont {U.}~\bibnamefont
  {Niemann}}, \bibinfo {author} {\bibfnamefont {Y.-M.}\ \bibnamefont {Wu}},
  \bibinfo {author} {\bibfnamefont {R.}~\bibnamefont {Oka}}, \bibinfo {author}
  {\bibfnamefont {D.}~\bibnamefont {Hirai}}, \bibinfo {author} {\bibfnamefont
  {Y.}~\bibnamefont {Wang}}, \bibinfo {author} {\bibfnamefont {Y.~E.}\
  \bibnamefont {Suyolcu}}, \bibinfo {author} {\bibfnamefont {M.}~\bibnamefont
  {Kim}}, \bibinfo {author} {\bibfnamefont {P.~A.}\ \bibnamefont {van Aken}},\
  and\ \bibinfo {author} {\bibfnamefont {H.}~\bibnamefont {Takagi}},\ }\href
  {https://doi.org/10.1073/pnas.2215722120} {\bibfield  {journal} {\bibinfo
  {journal} {Proc Natl Acad Sci U S A}\ }\textbf {\bibinfo {volume} {120}},\
  \bibinfo {pages} {e2215722120} (\bibinfo {year} {2023})}\BibitemShut
  {NoStop}%
\bibitem [{\citenamefont {Anisimov}\ \emph {et~al.}(1999)\citenamefont
  {Anisimov}, \citenamefont {Korotin}, \citenamefont {Z\"olfl}, \citenamefont
  {Pruschke}, \citenamefont {Le~Hur},\ and\ \citenamefont {Rice}}]{ani99}%
  \BibitemOpen
  \bibfield  {author} {\bibinfo {author} {\bibfnamefont {V.~I.}\ \bibnamefont
  {Anisimov}}, \bibinfo {author} {\bibfnamefont {M.~A.}\ \bibnamefont
  {Korotin}}, \bibinfo {author} {\bibfnamefont {M.}~\bibnamefont {Z\"olfl}},
  \bibinfo {author} {\bibfnamefont {T.}~\bibnamefont {Pruschke}}, \bibinfo
  {author} {\bibfnamefont {K.}~\bibnamefont {Le~Hur}},\ and\ \bibinfo {author}
  {\bibfnamefont {T.~M.}\ \bibnamefont {Rice}},\ }\href
  {https://doi.org/10.1103/PhysRevLett.83.364} {\bibfield  {journal} {\bibinfo
  {journal} {Phys. Rev. Lett.}\ }\textbf {\bibinfo {volume} {83}},\ \bibinfo
  {pages} {364} (\bibinfo {year} {1999})}\BibitemShut {NoStop}%
\bibitem [{\citenamefont {Hopkinson}\ and\ \citenamefont
  {Coleman}(2002)}]{hop02}%
  \BibitemOpen
  \bibfield  {author} {\bibinfo {author} {\bibfnamefont {J.}~\bibnamefont
  {Hopkinson}}\ and\ \bibinfo {author} {\bibfnamefont {P.}~\bibnamefont
  {Coleman}},\ }\href {https://doi.org/10.1103/PhysRevLett.89.267201}
  {\bibfield  {journal} {\bibinfo  {journal} {Phys. Rev. Lett.}\ }\textbf
  {\bibinfo {volume} {89}},\ \bibinfo {pages} {267201} (\bibinfo {year}
  {2002})}\BibitemShut {NoStop}%
\bibitem [{\citenamefont {Meyers}\ \emph {et~al.}(2014)\citenamefont {Meyers},
  \citenamefont {Middey}, \citenamefont {Cheng}, \citenamefont {Mukherjee},
  \citenamefont {Gray}, \citenamefont {Cao}, \citenamefont {Zhou},
  \citenamefont {Goodenough}, \citenamefont {Choi}, \citenamefont {Haskel},
  \citenamefont {Freeland}, \citenamefont {Saha-Dasgupta},\ and\ \citenamefont
  {Chakhalian}}]{mey14}%
  \BibitemOpen
  \bibfield  {author} {\bibinfo {author} {\bibfnamefont {D.}~\bibnamefont
  {Meyers}}, \bibinfo {author} {\bibfnamefont {S.}~\bibnamefont {Middey}},
  \bibinfo {author} {\bibfnamefont {J.~G.}\ \bibnamefont {Cheng}}, \bibinfo
  {author} {\bibfnamefont {S.}~\bibnamefont {Mukherjee}}, \bibinfo {author}
  {\bibfnamefont {B.~A.}\ \bibnamefont {Gray}}, \bibinfo {author}
  {\bibfnamefont {Y.}~\bibnamefont {Cao}}, \bibinfo {author} {\bibfnamefont
  {J.~S.}\ \bibnamefont {Zhou}}, \bibinfo {author} {\bibfnamefont {J.~B.}\
  \bibnamefont {Goodenough}}, \bibinfo {author} {\bibfnamefont
  {Y.}~\bibnamefont {Choi}}, \bibinfo {author} {\bibfnamefont {D.}~\bibnamefont
  {Haskel}}, \bibinfo {author} {\bibfnamefont {J.~W.}\ \bibnamefont
  {Freeland}}, \bibinfo {author} {\bibfnamefont {T.}~\bibnamefont
  {Saha-Dasgupta}},\ and\ \bibinfo {author} {\bibfnamefont {J.}~\bibnamefont
  {Chakhalian}},\ }\href {https://doi.org/10.1038/ncomms6818} {\bibfield
  {journal} {\bibinfo  {journal} {Nature Communications}\ }\textbf {\bibinfo
  {volume} {5}},\ \bibinfo {pages} {5818} (\bibinfo {year} {2014})}\BibitemShut
  {NoStop}%
\bibitem [{\citenamefont {Riegg}\ \emph {et~al.}(2016)\citenamefont {Riegg},
  \citenamefont {Widmann}, \citenamefont {Meir}, \citenamefont {Sterz},
  \citenamefont {G\"unther}, \citenamefont {B\"uttgen}, \citenamefont
  {Ebbinghaus}, \citenamefont {Reller}, \citenamefont {von Nidda},\ and\
  \citenamefont {Loidl}}]{rie16}%
  \BibitemOpen
  \bibfield  {author} {\bibinfo {author} {\bibfnamefont {S.}~\bibnamefont
  {Riegg}}, \bibinfo {author} {\bibfnamefont {S.}~\bibnamefont {Widmann}},
  \bibinfo {author} {\bibfnamefont {B.}~\bibnamefont {Meir}}, \bibinfo {author}
  {\bibfnamefont {S.}~\bibnamefont {Sterz}}, \bibinfo {author} {\bibfnamefont
  {A.}~\bibnamefont {G\"unther}}, \bibinfo {author} {\bibfnamefont
  {N.}~\bibnamefont {B\"uttgen}}, \bibinfo {author} {\bibfnamefont {S.~G.}\
  \bibnamefont {Ebbinghaus}}, \bibinfo {author} {\bibfnamefont
  {A.}~\bibnamefont {Reller}}, \bibinfo {author} {\bibfnamefont {H.-A.~K.}\
  \bibnamefont {von Nidda}},\ and\ \bibinfo {author} {\bibfnamefont
  {A.}~\bibnamefont {Loidl}},\ }\href
  {https://doi.org/10.1103/PhysRevB.93.115149} {\bibfield  {journal} {\bibinfo
  {journal} {Phys. Rev. B}\ }\textbf {\bibinfo {volume} {93}},\ \bibinfo
  {pages} {115149} (\bibinfo {year} {2016})}\BibitemShut {NoStop}%
\bibitem [{\citenamefont {Fulde}\ \emph {et~al.}(2001)\citenamefont {Fulde},
  \citenamefont {Yaresko}, \citenamefont {Zvyagin},\ and\ \citenamefont
  {Grin}}]{ful01}%
  \BibitemOpen
  \bibfield  {author} {\bibinfo {author} {\bibfnamefont {P.}~\bibnamefont
  {Fulde}}, \bibinfo {author} {\bibfnamefont {A.~N.}\ \bibnamefont {Yaresko}},
  \bibinfo {author} {\bibfnamefont {A.~A.}\ \bibnamefont {Zvyagin}},\ and\
  \bibinfo {author} {\bibfnamefont {Y.}~\bibnamefont {Grin}},\ }\href
  {https://doi.org/10.1209/epl/i2001-00322-3} {\bibfield  {journal} {\bibinfo
  {journal} {Europhysics Letters}\ }\textbf {\bibinfo {volume} {54}},\ \bibinfo
  {pages} {779} (\bibinfo {year} {2001})}\BibitemShut {NoStop}%
\bibitem [{\citenamefont {Arita}\ \emph {et~al.}(2007)\citenamefont {Arita},
  \citenamefont {Held}, \citenamefont {Lukoyanov},\ and\ \citenamefont
  {Anisimov}}]{ari04}%
  \BibitemOpen
  \bibfield  {author} {\bibinfo {author} {\bibfnamefont {R.}~\bibnamefont
  {Arita}}, \bibinfo {author} {\bibfnamefont {K.}~\bibnamefont {Held}},
  \bibinfo {author} {\bibfnamefont {A.~V.}\ \bibnamefont {Lukoyanov}},\ and\
  \bibinfo {author} {\bibfnamefont {V.~I.}\ \bibnamefont {Anisimov}},\ }\href
  {https://doi.org/10.1103/PhysRevLett.98.166402} {\bibfield  {journal}
  {\bibinfo  {journal} {Phys. Rev. Lett.}\ }\textbf {\bibinfo {volume} {98}},\
  \bibinfo {pages} {166402} (\bibinfo {year} {2007})}\BibitemShut {NoStop}%
\bibitem [{\citenamefont {J\"onsson}\ \emph {et~al.}(2007)\citenamefont
  {J\"onsson}, \citenamefont {Takenaka}, \citenamefont {Niitaka}, \citenamefont
  {Sasagawa}, \citenamefont {Sugai},\ and\ \citenamefont {Takagi}}]{joe07}%
  \BibitemOpen
  \bibfield  {author} {\bibinfo {author} {\bibfnamefont {P.~E.}\ \bibnamefont
  {J\"onsson}}, \bibinfo {author} {\bibfnamefont {K.}~\bibnamefont {Takenaka}},
  \bibinfo {author} {\bibfnamefont {S.}~\bibnamefont {Niitaka}}, \bibinfo
  {author} {\bibfnamefont {T.}~\bibnamefont {Sasagawa}}, \bibinfo {author}
  {\bibfnamefont {S.}~\bibnamefont {Sugai}},\ and\ \bibinfo {author}
  {\bibfnamefont {H.}~\bibnamefont {Takagi}},\ }\href
  {https://doi.org/10.1103/PhysRevLett.99.167402} {\bibfield  {journal}
  {\bibinfo  {journal} {Phys. Rev. Lett.}\ }\textbf {\bibinfo {volume} {99}},\
  \bibinfo {pages} {167402} (\bibinfo {year} {2007})}\BibitemShut {NoStop}%
\bibitem [{\citenamefont {Fujimori}\ \emph {et~al.}(1991)\citenamefont
  {Fujimori}, \citenamefont {Saeki},\ and\ \citenamefont {Nozaki}}]{fuj91}%
  \BibitemOpen
  \bibfield  {author} {\bibinfo {author} {\bibfnamefont {A.}~\bibnamefont
  {Fujimori}}, \bibinfo {author} {\bibfnamefont {M.}~\bibnamefont {Saeki}},\
  and\ \bibinfo {author} {\bibfnamefont {H.}~\bibnamefont {Nozaki}},\ }\href
  {https://doi.org/10.1103/PhysRevB.44.163} {\bibfield  {journal} {\bibinfo
  {journal} {Phys. Rev. B}\ }\textbf {\bibinfo {volume} {44}},\ \bibinfo
  {pages} {163} (\bibinfo {year} {1991})}\BibitemShut {NoStop}%
\bibitem [{\citenamefont {Suzuki}\ \emph {et~al.}(1993)\citenamefont {Suzuki},
  \citenamefont {Teshiam},\ and\ \citenamefont {Motizuki}}]{suz93}%
  \BibitemOpen
  \bibfield  {author} {\bibinfo {author} {\bibfnamefont {N.}~\bibnamefont
  {Suzuki}}, \bibinfo {author} {\bibfnamefont {T.}~\bibnamefont {Teshiam}},\
  and\ \bibinfo {author} {\bibfnamefont {K.}~\bibnamefont {Motizuki}},\ }\href
  {https://doi.org/10.7567/jjaps.32s3.299} {\bibfield  {journal} {\bibinfo
  {journal} {Japanese Journal of Applied Physics}\ }\textbf {\bibinfo {volume}
  {32}},\ \bibinfo {pages} {299} (\bibinfo {year} {1993})}\BibitemShut
  {NoStop}%
\bibitem [{\citenamefont {Knecht}\ \emph {et~al.}(1998)\citenamefont {Knecht},
  \citenamefont {Ebert},\ and\ \citenamefont {Bensch}}]{kne98}%
  \BibitemOpen
  \bibfield  {author} {\bibinfo {author} {\bibfnamefont {M.}~\bibnamefont
  {Knecht}}, \bibinfo {author} {\bibfnamefont {H.}~\bibnamefont {Ebert}},\ and\
  \bibinfo {author} {\bibfnamefont {W.}~\bibnamefont {Bensch}},\ }\href
  {https://doi.org/10.1088/0953-8984/10/42/011} {\bibfield  {journal} {\bibinfo
   {journal} {Journal of Physics: Condensed Matter}\ }\textbf {\bibinfo
  {volume} {10}},\ \bibinfo {pages} {9455} (\bibinfo {year}
  {1998})}\BibitemShut {NoStop}%
\bibitem [{\citenamefont {Koo}\ \emph {et~al.}(2001)\citenamefont {Koo},
  \citenamefont {Seo},\ and\ \citenamefont {Whangbo}}]{koo01}%
  \BibitemOpen
  \bibfield  {author} {\bibinfo {author} {\bibfnamefont {H.-J.}\ \bibnamefont
  {Koo}}, \bibinfo {author} {\bibfnamefont {D.-K.}\ \bibnamefont {Seo}},\ and\
  \bibinfo {author} {\bibfnamefont {M.-H.}\ \bibnamefont {Whangbo}},\ }\href
  {https://doi.org/https://doi.org/10.1006/jssc.2001.9260} {\bibfield
  {journal} {\bibinfo  {journal} {Journal of Solid State Chemistry}\ }\textbf
  {\bibinfo {volume} {160}},\ \bibinfo {pages} {287} (\bibinfo {year}
  {2001})}\BibitemShut {NoStop}%
\bibitem [{\citenamefont {Niu}\ \emph {et~al.}(2020)\citenamefont {Niu},
  \citenamefont {Zhang}, \citenamefont {Li}, \citenamefont {Yang},
  \citenamefont {Yan}, \citenamefont {Chen}, \citenamefont {Zhang},
  \citenamefont {Zhang}, \citenamefont {Ren}, \citenamefont {Gao},
  \citenamefont {Shi}, \citenamefont {Yu},\ and\ \citenamefont {Wu}}]{niu20}%
  \BibitemOpen
  \bibfield  {author} {\bibinfo {author} {\bibfnamefont {J.}~\bibnamefont
  {Niu}}, \bibinfo {author} {\bibfnamefont {W.}~\bibnamefont {Zhang}}, \bibinfo
  {author} {\bibfnamefont {Z.}~\bibnamefont {Li}}, \bibinfo {author}
  {\bibfnamefont {S.}~\bibnamefont {Yang}}, \bibinfo {author} {\bibfnamefont
  {D.}~\bibnamefont {Yan}}, \bibinfo {author} {\bibfnamefont {S.}~\bibnamefont
  {Chen}}, \bibinfo {author} {\bibfnamefont {Z.}~\bibnamefont {Zhang}},
  \bibinfo {author} {\bibfnamefont {Y.}~\bibnamefont {Zhang}}, \bibinfo
  {author} {\bibfnamefont {X.}~\bibnamefont {Ren}}, \bibinfo {author}
  {\bibfnamefont {P.}~\bibnamefont {Gao}}, \bibinfo {author} {\bibfnamefont
  {Y.}~\bibnamefont {Shi}}, \bibinfo {author} {\bibfnamefont {D.}~\bibnamefont
  {Yu}},\ and\ \bibinfo {author} {\bibfnamefont {X.}~\bibnamefont {Wu}},\
  }\href {https://doi.org/10.1088/1674-1056/abab85} {\bibfield  {journal}
  {\bibinfo  {journal} {Chinese Physics B}\ }\textbf {\bibinfo {volume} {29}},\
  \bibinfo {pages} {097104} (\bibinfo {year} {2020})}\BibitemShut {NoStop}%
\bibitem [{\citenamefont {Kawada}\ \emph {et~al.}(1975)\citenamefont {Kawada},
  \citenamefont {Nakano-Onoda}, \citenamefont {Ishii}, \citenamefont {Saeki},\
  and\ \citenamefont {Nakahira}}]{Kaw75}%
  \BibitemOpen
  \bibfield  {author} {\bibinfo {author} {\bibfnamefont {I.}~\bibnamefont
  {Kawada}}, \bibinfo {author} {\bibfnamefont {M.}~\bibnamefont
  {Nakano-Onoda}}, \bibinfo {author} {\bibfnamefont {M.}~\bibnamefont {Ishii}},
  \bibinfo {author} {\bibfnamefont {M.}~\bibnamefont {Saeki}},\ and\ \bibinfo
  {author} {\bibfnamefont {M.}~\bibnamefont {Nakahira}},\ }\href
  {https://doi.org/https://doi.org/10.1016/0022-4596(75)90209-1} {\bibfield
  {journal} {\bibinfo  {journal} {Journal of Solid State Chemistry}\ }\textbf
  {\bibinfo {volume} {15}},\ \bibinfo {pages} {246} (\bibinfo {year}
  {1975})}\BibitemShut {NoStop}%
\bibitem [{\citenamefont {Funahashi}\ \emph {et~al.}(1981)\citenamefont
  {Funahashi}, \citenamefont {Nozaki},\ and\ \citenamefont {Kawada}}]{fun81}%
  \BibitemOpen
  \bibfield  {author} {\bibinfo {author} {\bibfnamefont {S.}~\bibnamefont
  {Funahashi}}, \bibinfo {author} {\bibfnamefont {H.}~\bibnamefont {Nozaki}},\
  and\ \bibinfo {author} {\bibfnamefont {I.}~\bibnamefont {Kawada}},\ }\href
  {https://doi.org/https://doi.org/10.1016/0022-3697(81)90064-0} {\bibfield
  {journal} {\bibinfo  {journal} {Journal of Physics and Chemistry of Solids}\
  }\textbf {\bibinfo {volume} {42}},\ \bibinfo {pages} {1009 } (\bibinfo {year}
  {1981})}\BibitemShut {NoStop}%
\bibitem [{\citenamefont {Gauzzi}\ \emph {et~al.}(2019)\citenamefont {Gauzzi},
  \citenamefont {Moutaabbid}, \citenamefont {Klein}, \citenamefont {Loupias},\
  and\ \citenamefont {Hardy}}]{gau19}%
  \BibitemOpen
  \bibfield  {author} {\bibinfo {author} {\bibfnamefont {A.}~\bibnamefont
  {Gauzzi}}, \bibinfo {author} {\bibfnamefont {H.}~\bibnamefont {Moutaabbid}},
  \bibinfo {author} {\bibfnamefont {Y.}~\bibnamefont {Klein}}, \bibinfo
  {author} {\bibfnamefont {G.}~\bibnamefont {Loupias}},\ and\ \bibinfo {author}
  {\bibfnamefont {V.}~\bibnamefont {Hardy}},\ }\href
  {https://doi.org/10.1088/1361-648x/ab1d9b} {\bibfield  {journal} {\bibinfo
  {journal} {Journal of Physics: Condensed Matter}\ }\textbf {\bibinfo {volume}
  {31}},\ \bibinfo {pages} {31LT01} (\bibinfo {year} {2019})}\BibitemShut
  {NoStop}%
\bibitem [{\citenamefont {Barua}\ \emph {et~al.}(2017)\citenamefont {Barua},
  \citenamefont {Hatnean}, \citenamefont {Lees},\ and\ \citenamefont
  {Balakrishnan}}]{bar17}%
  \BibitemOpen
  \bibfield  {author} {\bibinfo {author} {\bibfnamefont {S.}~\bibnamefont
  {Barua}}, \bibinfo {author} {\bibfnamefont {M.~C.}\ \bibnamefont {Hatnean}},
  \bibinfo {author} {\bibfnamefont {M.~R.}\ \bibnamefont {Lees}},\ and\
  \bibinfo {author} {\bibfnamefont {G.}~\bibnamefont {Balakrishnan}},\ }\href
  {https://doi.org/10.1038/s41598-017-11247-4} {\bibfield  {journal} {\bibinfo
  {journal} {Scientific Reports}\ }\textbf {\bibinfo {volume} {7}},\ \bibinfo
  {pages} {10964} (\bibinfo {year} {2017})}\BibitemShut {NoStop}%
\bibitem [{\citenamefont {Saeki}\ \emph {et~al.}(1974)\citenamefont {Saeki},
  \citenamefont {Nakano},\ and\ \citenamefont {Nakahira}}]{sae74}%
  \BibitemOpen
  \bibfield  {author} {\bibinfo {author} {\bibfnamefont {M.}~\bibnamefont
  {Saeki}}, \bibinfo {author} {\bibfnamefont {M.}~\bibnamefont {Nakano}},\ and\
  \bibinfo {author} {\bibfnamefont {M.}~\bibnamefont {Nakahira}},\ }\href
  {https://doi.org/https://doi.org/10.1016/0022-0248(74)90294-2} {\bibfield
  {journal} {\bibinfo  {journal} {Journal of Crystal Growth}\ }\textbf
  {\bibinfo {volume} {24-25}},\ \bibinfo {pages} {154} (\bibinfo {year}
  {1974})}\BibitemShut {NoStop}%
\bibitem [{\citenamefont {Nishihara}\ \emph {et~al.}(1977)\citenamefont
  {Nishihara}, \citenamefont {Yasuoka}, \citenamefont {Oka}, \citenamefont
  {Kosuge},\ and\ \citenamefont {Kachi}}]{nis77}%
  \BibitemOpen
  \bibfield  {author} {\bibinfo {author} {\bibfnamefont {H.}~\bibnamefont
  {Nishihara}}, \bibinfo {author} {\bibfnamefont {H.}~\bibnamefont {Yasuoka}},
  \bibinfo {author} {\bibfnamefont {Y.}~\bibnamefont {Oka}}, \bibinfo {author}
  {\bibfnamefont {K.}~\bibnamefont {Kosuge}},\ and\ \bibinfo {author}
  {\bibfnamefont {S.}~\bibnamefont {Kachi}},\ }\href
  {https://doi.org/10.1143/JPSJ.42.787} {\bibfield  {journal} {\bibinfo
  {journal} {Journal of the Physical Society of Japan}\ }\textbf {\bibinfo
  {volume} {42}},\ \bibinfo {pages} {787} (\bibinfo {year} {1977})}\BibitemShut
  {NoStop}%
\bibitem [{\citenamefont {Inoue}\ \emph {et~al.}(1986)\citenamefont {Inoue},
  \citenamefont {Muneta}, \citenamefont {Negishi},\ and\ \citenamefont
  {Sasaki}}]{ino86}%
  \BibitemOpen
  \bibfield  {author} {\bibinfo {author} {\bibfnamefont {M.}~\bibnamefont
  {Inoue}}, \bibinfo {author} {\bibfnamefont {Y.}~\bibnamefont {Muneta}},
  \bibinfo {author} {\bibfnamefont {H.}~\bibnamefont {Negishi}},\ and\ \bibinfo
  {author} {\bibfnamefont {M.}~\bibnamefont {Sasaki}},\ }\href
  {https://doi.org/10.1007/BF00683766} {\bibfield  {journal} {\bibinfo
  {journal} {Journal of Low Temperature Physics}\ }\textbf {\bibinfo {volume}
  {63}},\ \bibinfo {pages} {235} (\bibinfo {year} {1986})}\BibitemShut
  {NoStop}%
\bibitem [{\citenamefont {Maeno}\ \emph {et~al.}(1997)\citenamefont {Maeno},
  \citenamefont {Yoshida}, \citenamefont {Hashimoto}, \citenamefont
  {Nishizaki}, \citenamefont {Ikeda}, \citenamefont {Nohara}, \citenamefont
  {Fujita}, \citenamefont {Mackenzie}, \citenamefont {Hussey}, \citenamefont
  {Bednorz},\ and\ \citenamefont {Lichtenberg}}]{mae97}%
  \BibitemOpen
  \bibfield  {author} {\bibinfo {author} {\bibfnamefont {Y.}~\bibnamefont
  {Maeno}}, \bibinfo {author} {\bibfnamefont {K.}~\bibnamefont {Yoshida}},
  \bibinfo {author} {\bibfnamefont {H.}~\bibnamefont {Hashimoto}}, \bibinfo
  {author} {\bibfnamefont {S.}~\bibnamefont {Nishizaki}}, \bibinfo {author}
  {\bibfnamefont {S.-i.}\ \bibnamefont {Ikeda}}, \bibinfo {author}
  {\bibfnamefont {M.}~\bibnamefont {Nohara}}, \bibinfo {author} {\bibfnamefont
  {T.}~\bibnamefont {Fujita}}, \bibinfo {author} {\bibfnamefont {A.~P.}\
  \bibnamefont {Mackenzie}}, \bibinfo {author} {\bibfnamefont {N.~E.}\
  \bibnamefont {Hussey}}, \bibinfo {author} {\bibfnamefont {J.~G.}\
  \bibnamefont {Bednorz}},\ and\ \bibinfo {author} {\bibfnamefont
  {F.}~\bibnamefont {Lichtenberg}},\ }\href
  {https://doi.org/10.1143/JPSJ.66.1405} {\bibfield  {journal} {\bibinfo
  {journal} {Journal of the Physical Society of Japan}\ }\textbf {\bibinfo
  {volume} {66}},\ \bibinfo {pages} {1405} (\bibinfo {year}
  {1997})}\BibitemShut {NoStop}%
\bibitem [{\citenamefont {Johnston}\ \emph {et~al.}(1999)\citenamefont
  {Johnston}, \citenamefont {Swenson},\ and\ \citenamefont {Kondo}}]{joh99}%
  \BibitemOpen
  \bibfield  {author} {\bibinfo {author} {\bibfnamefont {D.~C.}\ \bibnamefont
  {Johnston}}, \bibinfo {author} {\bibfnamefont {C.~A.}\ \bibnamefont
  {Swenson}},\ and\ \bibinfo {author} {\bibfnamefont {S.}~\bibnamefont
  {Kondo}},\ }\href {https://doi.org/10.1103/PhysRevB.59.2627} {\bibfield
  {journal} {\bibinfo  {journal} {Phys. Rev. B}\ }\textbf {\bibinfo {volume}
  {59}},\ \bibinfo {pages} {2627} (\bibinfo {year} {1999})}\BibitemShut
  {NoStop}%
\bibitem [{\citenamefont {Limelette}\ \emph {et~al.}(2005)\citenamefont
  {Limelette}, \citenamefont {Hardy}, \citenamefont {Auban-Senzier},
  \citenamefont {J\'erome}, \citenamefont {Flahaut}, \citenamefont {H\'ebert},
  \citenamefont {Fr\'esard}, \citenamefont {Simon}, \citenamefont {Noudem},\
  and\ \citenamefont {Maignan}}]{lim05}%
  \BibitemOpen
  \bibfield  {author} {\bibinfo {author} {\bibfnamefont {P.}~\bibnamefont
  {Limelette}}, \bibinfo {author} {\bibfnamefont {V.}~\bibnamefont {Hardy}},
  \bibinfo {author} {\bibfnamefont {P.}~\bibnamefont {Auban-Senzier}}, \bibinfo
  {author} {\bibfnamefont {D.}~\bibnamefont {J\'erome}}, \bibinfo {author}
  {\bibfnamefont {D.}~\bibnamefont {Flahaut}}, \bibinfo {author} {\bibfnamefont
  {S.}~\bibnamefont {H\'ebert}}, \bibinfo {author} {\bibfnamefont
  {R.}~\bibnamefont {Fr\'esard}}, \bibinfo {author} {\bibfnamefont
  {C.}~\bibnamefont {Simon}}, \bibinfo {author} {\bibfnamefont
  {J.}~\bibnamefont {Noudem}},\ and\ \bibinfo {author} {\bibfnamefont
  {A.}~\bibnamefont {Maignan}},\ }\href
  {https://doi.org/10.1103/PhysRevB.71.233108} {\bibfield  {journal} {\bibinfo
  {journal} {Phys. Rev. B}\ }\textbf {\bibinfo {volume} {71}},\ \bibinfo
  {pages} {233108} (\bibinfo {year} {2005})}\BibitemShut {NoStop}%
\bibitem [{\citenamefont {Hellwege}\ and\ \citenamefont
  {Hellwege}(1976)}]{hel76}%
  \BibitemOpen
  \bibfield  {author} {\bibinfo {author} {\bibfnamefont {K.~H.}\ \bibnamefont
  {Hellwege}}\ and\ \bibinfo {author} {\bibfnamefont {A.~M.}\ \bibnamefont
  {Hellwege}},\ }\href@noop {} {\emph {\bibinfo {title} {Landolt-B{\"o}rnstein,
  Group II}}},\ Vol.~\bibinfo {volume} {8}\ (\bibinfo  {publisher}
  {Springer-Verlag},\ \bibinfo {address} {Berlin, Germany},\ \bibinfo {year}
  {1976})\ p.~\bibinfo {pages} {27}\BibitemShut {NoStop}%
\bibitem [{\citenamefont {Nozaki}\ \emph {et~al.}(1978)\citenamefont {Nozaki},
  \citenamefont {Umehara}, \citenamefont {Ishizawa}, \citenamefont {Saeki},
  \citenamefont {Mizoguchi},\ and\ \citenamefont {Nakahira}}]{noz78}%
  \BibitemOpen
  \bibfield  {author} {\bibinfo {author} {\bibfnamefont {H.}~\bibnamefont
  {Nozaki}}, \bibinfo {author} {\bibfnamefont {M.}~\bibnamefont {Umehara}},
  \bibinfo {author} {\bibfnamefont {Y.}~\bibnamefont {Ishizawa}}, \bibinfo
  {author} {\bibfnamefont {M.}~\bibnamefont {Saeki}}, \bibinfo {author}
  {\bibfnamefont {T.}~\bibnamefont {Mizoguchi}},\ and\ \bibinfo {author}
  {\bibfnamefont {M.}~\bibnamefont {Nakahira}},\ }\href
  {https://doi.org/https://doi.org/10.1016/0022-3697(78)90144-0} {\bibfield
  {journal} {\bibinfo  {journal} {Journal of Physics and Chemistry of Solids}\
  }\textbf {\bibinfo {volume} {39}},\ \bibinfo {pages} {851 } (\bibinfo {year}
  {1978})}\BibitemShut {NoStop}%
\bibitem [{\citenamefont {Nakanishi}\ \emph {et~al.}(2000)\citenamefont
  {Nakanishi}, \citenamefont {Yoshimura}, \citenamefont {Kosuge}, \citenamefont
  {Goto}, \citenamefont {Fujii},\ and\ \citenamefont {Takada}}]{nak00}%
  \BibitemOpen
  \bibfield  {author} {\bibinfo {author} {\bibfnamefont {M.}~\bibnamefont
  {Nakanishi}}, \bibinfo {author} {\bibfnamefont {K.}~\bibnamefont
  {Yoshimura}}, \bibinfo {author} {\bibfnamefont {K.}~\bibnamefont {Kosuge}},
  \bibinfo {author} {\bibfnamefont {T.}~\bibnamefont {Goto}}, \bibinfo {author}
  {\bibfnamefont {T.}~\bibnamefont {Fujii}},\ and\ \bibinfo {author}
  {\bibfnamefont {J.}~\bibnamefont {Takada}},\ }\href
  {https://doi.org/https://doi.org/10.1016/S0304-8853(00)00509-6} {\bibfield
  {journal} {\bibinfo  {journal} {Journal of Magnetism and Magnetic Materials}\
  }\textbf {\bibinfo {volume} {221}},\ \bibinfo {pages} {301} (\bibinfo {year}
  {2000})}\BibitemShut {NoStop}%
\bibitem [{\citenamefont {Lee}\ \emph {et~al.}(1986)\citenamefont {Lee},
  \citenamefont {Rice}, \citenamefont {Serene}, \citenamefont {Sham},\ and\
  \citenamefont {Wilkins}}]{lee86}%
  \BibitemOpen
  \bibfield  {author} {\bibinfo {author} {\bibfnamefont {P.}~\bibnamefont
  {Lee}}, \bibinfo {author} {\bibfnamefont {T.}~\bibnamefont {Rice}}, \bibinfo
  {author} {\bibfnamefont {J.}~\bibnamefont {Serene}}, \bibinfo {author}
  {\bibfnamefont {L.}~\bibnamefont {Sham}},\ and\ \bibinfo {author}
  {\bibfnamefont {J.}~\bibnamefont {Wilkins}},\ }\href
  {http://inis.iaea.org/search/search.aspx?orig_q=RN:17058516} {\bibfield
  {journal} {\bibinfo  {journal} {Comments on Condensed Matter Physics}\
  }\textbf {\bibinfo {volume} {12}},\ \bibinfo {pages} {99} (\bibinfo {year}
  {1986})}\BibitemShut {NoStop}%
\bibitem [{\citenamefont {Wilson}(1975)}]{wil75}%
  \BibitemOpen
  \bibfield  {author} {\bibinfo {author} {\bibfnamefont {K.~G.}\ \bibnamefont
  {Wilson}},\ }\href {https://doi.org/10.1103/RevModPhys.47.773} {\bibfield
  {journal} {\bibinfo  {journal} {Rev. Mod. Phys.}\ }\textbf {\bibinfo {volume}
  {47}},\ \bibinfo {pages} {773} (\bibinfo {year} {1975})}\BibitemShut
  {NoStop}%
\bibitem [{\citenamefont {{Nozi\`eres, Ph.}}\ and\ \citenamefont {{Blandin,
  A.}}(1980)}]{noz80}%
  \BibitemOpen
  \bibfield  {author} {\bibinfo {author} {\bibnamefont {{Nozi\`eres, Ph.}}}\
  and\ \bibinfo {author} {\bibnamefont {{Blandin, A.}}},\ }\href
  {https://doi.org/10.1051/jphys:01980004103019300} {\bibfield  {journal}
  {\bibinfo  {journal} {J. Phys. France}\ }\textbf {\bibinfo {volume} {41}},\
  \bibinfo {pages} {193} (\bibinfo {year} {1980})}\BibitemShut {NoStop}%
\bibitem [{\citenamefont {Nozaki}\ \emph {et~al.}(1975)\citenamefont {Nozaki},
  \citenamefont {Ishizawa}, \citenamefont {Saeki},\ and\ \citenamefont
  {Nakahira}}]{noz75}%
  \BibitemOpen
  \bibfield  {author} {\bibinfo {author} {\bibfnamefont {H.}~\bibnamefont
  {Nozaki}}, \bibinfo {author} {\bibfnamefont {Y.}~\bibnamefont {Ishizawa}},
  \bibinfo {author} {\bibfnamefont {M.}~\bibnamefont {Saeki}},\ and\ \bibinfo
  {author} {\bibfnamefont {M.}~\bibnamefont {Nakahira}},\ }\href
  {https://doi.org/https://doi.org/10.1016/0375-9601(75)90593-9} {\bibfield
  {journal} {\bibinfo  {journal} {Physics Letters A}\ }\textbf {\bibinfo
  {volume} {54}},\ \bibinfo {pages} {29 } (\bibinfo {year} {1975})}\BibitemShut
  {NoStop}%
\bibitem [{\citenamefont {Knebel}\ \emph {et~al.}(1997)\citenamefont {Knebel},
  \citenamefont {Eggert}, \citenamefont {Schmidt}, \citenamefont {Krimmel},
  \citenamefont {Dressel},\ and\ \citenamefont {Loidl}}]{kne97}%
  \BibitemOpen
  \bibfield  {author} {\bibinfo {author} {\bibfnamefont {G.}~\bibnamefont
  {Knebel}}, \bibinfo {author} {\bibfnamefont {C.}~\bibnamefont {Eggert}},
  \bibinfo {author} {\bibfnamefont {T.}~\bibnamefont {Schmidt}}, \bibinfo
  {author} {\bibfnamefont {A.}~\bibnamefont {Krimmel}}, \bibinfo {author}
  {\bibfnamefont {M.}~\bibnamefont {Dressel}},\ and\ \bibinfo {author}
  {\bibfnamefont {A.}~\bibnamefont {Loidl}},\ }\href
  {https://doi.org/https://doi.org/10.1016/S0921-4526(96)00756-9} {\bibfield
  {journal} {\bibinfo  {journal} {Physica B: Condensed Matter}\ }\textbf
  {\bibinfo {volume} {230-232}},\ \bibinfo {pages} {593} (\bibinfo {year}
  {1997})}\BibitemShut {NoStop}%
\bibitem [{\citenamefont {Lee}\ \emph {et~al.}(2010)\citenamefont {Lee},
  \citenamefont {Kurita}, \citenamefont {chun Ho}, \citenamefont {Condron},
  \citenamefont {Klavins}, \citenamefont {Kauzlarich}, \citenamefont {Maple},
  \citenamefont {Movshovich}, \citenamefont {Bauer}, \citenamefont {Thompson},\
  and\ \citenamefont {Fisk}}]{lee10}%
  \BibitemOpen
  \bibfield  {author} {\bibinfo {author} {\bibfnamefont {H.-O.}\ \bibnamefont
  {Lee}}, \bibinfo {author} {\bibfnamefont {N.}~\bibnamefont {Kurita}},
  \bibinfo {author} {\bibfnamefont {P.}~\bibnamefont {chun Ho}}, \bibinfo
  {author} {\bibfnamefont {C.~L.}\ \bibnamefont {Condron}}, \bibinfo {author}
  {\bibfnamefont {P.}~\bibnamefont {Klavins}}, \bibinfo {author} {\bibfnamefont
  {S.~M.}\ \bibnamefont {Kauzlarich}}, \bibinfo {author} {\bibfnamefont
  {M.~B.}\ \bibnamefont {Maple}}, \bibinfo {author} {\bibfnamefont
  {R.}~\bibnamefont {Movshovich}}, \bibinfo {author} {\bibfnamefont {E.~D.}\
  \bibnamefont {Bauer}}, \bibinfo {author} {\bibfnamefont {J.~D.}\ \bibnamefont
  {Thompson}},\ and\ \bibinfo {author} {\bibfnamefont {Z.}~\bibnamefont
  {Fisk}},\ }\href {https://doi.org/10.1088/0953-8984/22/6/065601} {\bibfield
  {journal} {\bibinfo  {journal} {Journal of Physics: Condensed Matter}\
  }\textbf {\bibinfo {volume} {22}},\ \bibinfo {pages} {065601} (\bibinfo
  {year} {2010})}\BibitemShut {NoStop}%
\bibitem [{\citenamefont {Hossain}\ \emph {et~al.}(2000)\citenamefont
  {Hossain}, \citenamefont {Hamashima}, \citenamefont {Umeo}, \citenamefont
  {Takabatake}, \citenamefont {Geibel},\ and\ \citenamefont
  {Steglich}}]{hos00}%
  \BibitemOpen
  \bibfield  {author} {\bibinfo {author} {\bibfnamefont {Z.}~\bibnamefont
  {Hossain}}, \bibinfo {author} {\bibfnamefont {S.}~\bibnamefont {Hamashima}},
  \bibinfo {author} {\bibfnamefont {K.}~\bibnamefont {Umeo}}, \bibinfo {author}
  {\bibfnamefont {T.}~\bibnamefont {Takabatake}}, \bibinfo {author}
  {\bibfnamefont {C.}~\bibnamefont {Geibel}},\ and\ \bibinfo {author}
  {\bibfnamefont {F.}~\bibnamefont {Steglich}},\ }\href
  {https://doi.org/10.1103/PhysRevB.62.8950} {\bibfield  {journal} {\bibinfo
  {journal} {Phys. Rev. B}\ }\textbf {\bibinfo {volume} {62}},\ \bibinfo
  {pages} {8950} (\bibinfo {year} {2000})}\BibitemShut {NoStop}%
\bibitem [{\citenamefont {McWhan}\ and\ \citenamefont {Rice}(1969)}]{mcw69}%
  \BibitemOpen
  \bibfield  {author} {\bibinfo {author} {\bibfnamefont {D.~B.}\ \bibnamefont
  {McWhan}}\ and\ \bibinfo {author} {\bibfnamefont {T.~M.}\ \bibnamefont
  {Rice}},\ }\href {https://doi.org/10.1103/PhysRevLett.22.887} {\bibfield
  {journal} {\bibinfo  {journal} {Phys. Rev. Lett.}\ }\textbf {\bibinfo
  {volume} {22}},\ \bibinfo {pages} {887} (\bibinfo {year} {1969})}\BibitemShut
  {NoStop}%
\bibitem [{\citenamefont {Kadowaki}\ and\ \citenamefont {Woods}(1986)}]{kad86}%
  \BibitemOpen
  \bibfield  {author} {\bibinfo {author} {\bibfnamefont {K.}~\bibnamefont
  {Kadowaki}}\ and\ \bibinfo {author} {\bibfnamefont {S.~B.}\ \bibnamefont
  {Woods}},\ }\href {https://doi.org/10.1016/0038-1098(86)90785-4} {\bibfield
  {journal} {\bibinfo  {journal} {Solid State Communications}\ }\textbf
  {\bibinfo {volume} {58}},\ \bibinfo {pages} {507} (\bibinfo {year} {1986})},\
  \bibinfo {note} {00787}\BibitemShut {NoStop}%
\bibitem [{\citenamefont {Nicklas}(2015)}]{Nic15}%
  \BibitemOpen
  \bibfield  {author} {\bibinfo {author} {\bibfnamefont {M.}~\bibnamefont
  {Nicklas}},\ }in\ \href {https://doi.org/10.1007/978-3-662-44133-6} {\emph
  {\bibinfo {booktitle} {Strongly Correlated Systems: Experimental
  Techniques}}},\ \bibinfo {series} {Collection of modern experimental methods
  for strongly correlated systems}, Vol.\ \bibinfo {volume} {180}\ (\bibinfo
  {publisher} {Springer},\ \bibinfo {address} {Berlin},\ \bibinfo {year}
  {2015})\ Chap.~\bibinfo {chapter} {6}, pp.\ \bibinfo {pages}
  {173--204}\BibitemShut {NoStop}%
\end{thebibliography}%

\end{document}